\documentclass{aa}
\usepackage{graphics}
\def\a{{$\alpha$}}

\def\gsnr{{G~116.5+1.1}}
\def\newg{{G~117.4+1.5}}
\newcommand{\h}{$^{\rm h}$}
\newcommand{\m}{$^{\rm m}$}
\newcommand{\s}{$^{\rm s}$}
\newcommand{\dd}{$\delta$}
\newcommand{\ha}{\rm H$\alpha$}
\newcommand{\hbeta}{\rm H$\beta$}
\newcommand{\HII}{\ion{H}{ii}}

\newcommand{\hnii}{{\rm H}$\alpha+[$\ion{N}{ii}$]$}
\newcommand{\nii}{$[$\ion{N}{ii}$]$}
\newcommand{\sii}{$[$\ion{S}{ii}$]$}
\newcommand{\oi}{$[$\ion{O}{i}$]$}

\newcommand{\oiii}{$[$\ion{O}{iii}$]$}
\newcommand{\snr}{\rm supernova remnant}
\newcommand{\snrs}{\rm supernova remnants}
\newcommand{\et}{et al.}
\newcommand{\flux}{$10^{-17}$ erg s$^{-1}$ cm$^{-2}$ arcsec$^{-2}$}
\newcommand{\dens}{\rm cm$^{-3}$}
\newcommand{\sdens}{\rm cm$^{-2}$}
\newcommand{\vel}{\rm km s$^{-1}$}
\newcommand{\mm}{\rm $\mu$m}
\begin{document}

%
\title{The faint supernova remnant \gsnr\ and the detection of a new 
candidate remnant}
\author{F. Mavromatakis\inst{1},
 P. Boumis \inst{2},
 E. Xilouris \inst{2},
 J. Papamastorakis\inst{1,3},
and J. Alikakos\inst{2,4}}
\offprints{F. Mavromatakis,\email{fotis@physics.uoc.gr}}
\authorrunning{F. Mavromatakis}
\titlerunning{The supernova remnants \gsnr\ and \newg}
\institute{
University of Crete, Physics Department, P.O. Box 2208, 710 03 Heraklion, 
Crete, Greece
\and
Institute of Astronomy \& Astrophysics, National Observatory of Athens,
I. Metaxa \& V. Pavlou, P. Penteli, 15236 Athens, Greece 
\and
Foundation for Research and Technology-Hellas, P.O. Box 1527, 711 10 
Heraklion, Crete, Greece
\and
Astronomical Laboratory, Department of Physics, University of Patras, 26500
Rio--Patras, Greece
}
\date{Received 15 October 2004/Accepted 13 January 2005}

\abstract{
The extended \snr\ \gsnr\ was observed in the optical emission lines
of \hnii,
\sii\ and \oiii; deep long--slit spectra were also obtained. The
morphology of the remnant's observed emission is mainly diffuse and
patchy in contrast to the known filamentary emission seen along the
western limb. The bulk of the detected emission in the region appears
unrelated to the remnant but there is one area of emission in the
south--east which is characterized by a \sii/\ha\ ratio of $\sim$0.5,
implying a possible relation to \gsnr. If this is actually the case,
it would imply a more extended remnant than previously realized.
Emission in the \oiii\ 5007 \AA\ line image is not detected, excluding
moderate or fast velocity shocks running into ionized interstellar
clouds. Our current estimate of the distance to \gsnr\ of $\sim$3 kpc
is in agreement with earlier estimates and implies a very extended
remnant (69 pc $\times$ 45 pc).
\par
Observations further to the north--east of \gsnr\ revealed a network
of filamentary structures prominent in
\hnii\ and \sii\ but failed to detect \oiii\ line emission. 
Long-slit spectra in a number of positions provide strong evidence
that this newly detected emission arises from shock heated
gas. Typical \ha\ fluxes lie in the range of 9 to 17 $\cdot$ \flux,
while low electron densities are implied by the intensities of the
sulfur lines. Weak emission from the medium ionization line at 5007
\AA\ is detected in only one spectrum. Cool dust emission at 60 and
100 microns may be correlated with the optical emission in a limited
number of positions. Surpisingly, radio emission is not detected in
published surveys suggesting that the new candidate remnant may belong
to the class of ``radio quiet'' supernova remnants.
\keywords{ISM: general -- ISM: supernova remnants
-- ISM: individual objects: G 116.5+1.1.}
}
\maketitle
\section{Introduction}
Supernova remnants are important ingredients of a galaxy because of the huge 
amount of energy released during the initial explosion, while in subsequent 
phases heavy elements are mixed into the interstellar medium (ISM). 
Observations of \snrs\ in X--ray wavelengths allow us to directly probe the 
hot gas
inside the primary shock wave. Optical observations offer an important tool for the study of the
interaction of the shock wave with dense concentrations of gas found in the
ISM. The detection and observation of well extended remnants is 
a difficult task since their surface brightness can be quite low but 
such observations
would contribute significantly towards an unbiased \snr\ catalog.
\par
One such extended object is \object{G 116.5+1.1} located in the
Perseus Arm. Reich \& Braunsfurth (\cite{rei81}) used data at 2.7 and 1.4 GHz to
establish the non--thermal nature of the radio emission and classified \gsnr\
as a \snr. Identifying certain HI features as related to the object, they
obtained a  distance of 4.4 ($\pm$0.4) kpc. Fich (\cite{fic86}) in a
study of large--scale structures in the Perseus Arm suggested that \gsnr\ as
well as CTB~1 are found at the edges of an HI supershell. 
Reich \& Braunsfurth (\cite{rei81}) provided quantitative estimates
of the remnant's explosion energy, shock velocity, ISM density 
etc., but with significant uncertainties. Unfortunately, there 
have not been any dedicated X--ray observations, while no emission was 
detected in the ROSAT All--Sky survey.
\begin{table}
      \caption[]{Imaging log}
         \label{filters}
\begin{flushleft}
\begin{tabular}{lllllll}
            \noalign{\smallskip}
\hline
  \hnii\        & \sii\         & \oiii\ & Date &   \cr
\hline
2400$^{\rm a}$(1)$^{\rm b}$   &4800 (2)&--   & 09-08-2002 &   \cr
 \hline
2400(1)                      &2400 (1)&--    & 10-08-2002 &   \cr
 \hline
--                           & --     &7200(3)& 11-08-2002 &   \cr
 \hline
2400(1)                      &2400 (1)&--    & 12-08-2002 &   \cr
 \hline
4800(2) 	             &4800 (2)&--    & 13-08-2002 &   \cr
\hline
\hline
4800(2)$^{*}$ 	             &--      &--    & 14-06-2004 &   \cr
 \hline
4800(2)$^{*}$ 	             &--      &--    & 15-06-2004 &   \cr
 \hline
--          	             &2400 (1)&--    & 20-06-2004 &   \cr
 \hline
2400(1) 	             &--      &--    & 21-06-2004 &   \cr
 \hline
2400(1) 	             &2400 (1)&2400 (1)& 22-06-2004 &   \cr
 \hline

 \hline
6560$^{\rm c}$(75$^{\rm d}$)  & 6708 (20) & 5005 (28 ) & \cr
 \hline
\end{tabular}
\end{flushleft}
${\rm ^a}$ Total exposure time in sec \\\
${\rm ^b}$ Number of individual frames \\\
${\rm ^c}$ Central wavelength in \AA\  \\\
${\rm ^d}$ Full width at half maximum in \AA\ \\\
$^{*}$ Each of the two frames covers different fields
   \end{table}

  \begin{table}
      \caption[]{Spectral log}
         \label{spectra}
\begin{flushleft}
\begin{tabular}{lllll}
            \noalign{\smallskip}
\hline
	Slit centers &  & Exp. times$^{\rm a}$ &  Position$^{\rm c}$ \cr
\hline
   $\alpha$ & $\delta$                     & ($\#$)$^{\rm b}$  &  \cr
 \hline 
23\h49\m09\s & 63\degr30\arcmin27\arcsec 	& 7800 (2)  & I  \cr
 \hline
23\h49\m18\s & 63\degr33\arcmin06\arcsec 	& 3600 (1)  & II \cr
 \hline
23\h55\m45\s & 62\degr53\arcmin12\arcsec 	& 7200 (2)  & III \cr
\hline
23\h58\m00\s & 63\degr23\arcmin00\arcsec 	& 7800 (2)  & IV \cr
 \hline
\hline
00\h00\m38\s & 63\degr16\arcmin03\arcsec        & 7200 (2)  & V \cr
\hline
00\h02\m55\s & 63\degr17\arcmin59\arcsec        & 7200 (2)  & VI \cr 
\hline
00\h02\m32\s & 63\degr30\arcmin04\arcsec        & 7200 (2)  & VII \cr
\hline
23\h58\m22\s & 63\degr55\arcmin14\arcsec        & 3600 (1)  & VIII \cr
\hline
\end{tabular}
\end{flushleft}
${\rm ^a}$ Total exposure time in sec\\\
${\rm ^b}$ Number of spectra acquired \\\
${\rm ^c}$ Marked as in Fig. \ref{fig01} \\\
   \end{table}
\par
Fesen \et\ (\cite{fes97}) obtained
\ha\ and \sii\ images (59\arcmin$\times$52\arcmin\  wide) in the north--west 
and detected a faint long ($\sim$30\arcmin) filament. 
Low dispersion spectra at the
position of the filament verified the shock heated nature of this
optical emission. Because our knowledge of this remnant is limited to optical 
wavelengths we obtained the deepest currently available
 CCD images of the full field 
of \gsnr\ in the emission lines of \hnii, \sii\ and \oiii. The detection of
new filamentary structures to the north--east of \gsnr\ led to additional 
imagery in the same filters.
Deep long--slit spectra were also acquired at various locations in 
order to study in more detail the new structures.
Information about the observations and the data reduction is given 
in Sect. 2. In Sect. 3 and 4 the results of the imaging and spectral 
observations are presented, while in Sect. 5 we report on observations  
in wavelengths other than optical. In Sect. 6 we discuss the properties 
of \gsnr\ and the new candidate remnant. In Sect. 7 we summarize the 
results of this work.
\section{Observations}
  \begin {figure*}
  \resizebox{\hsize}{!}{\includegraphics{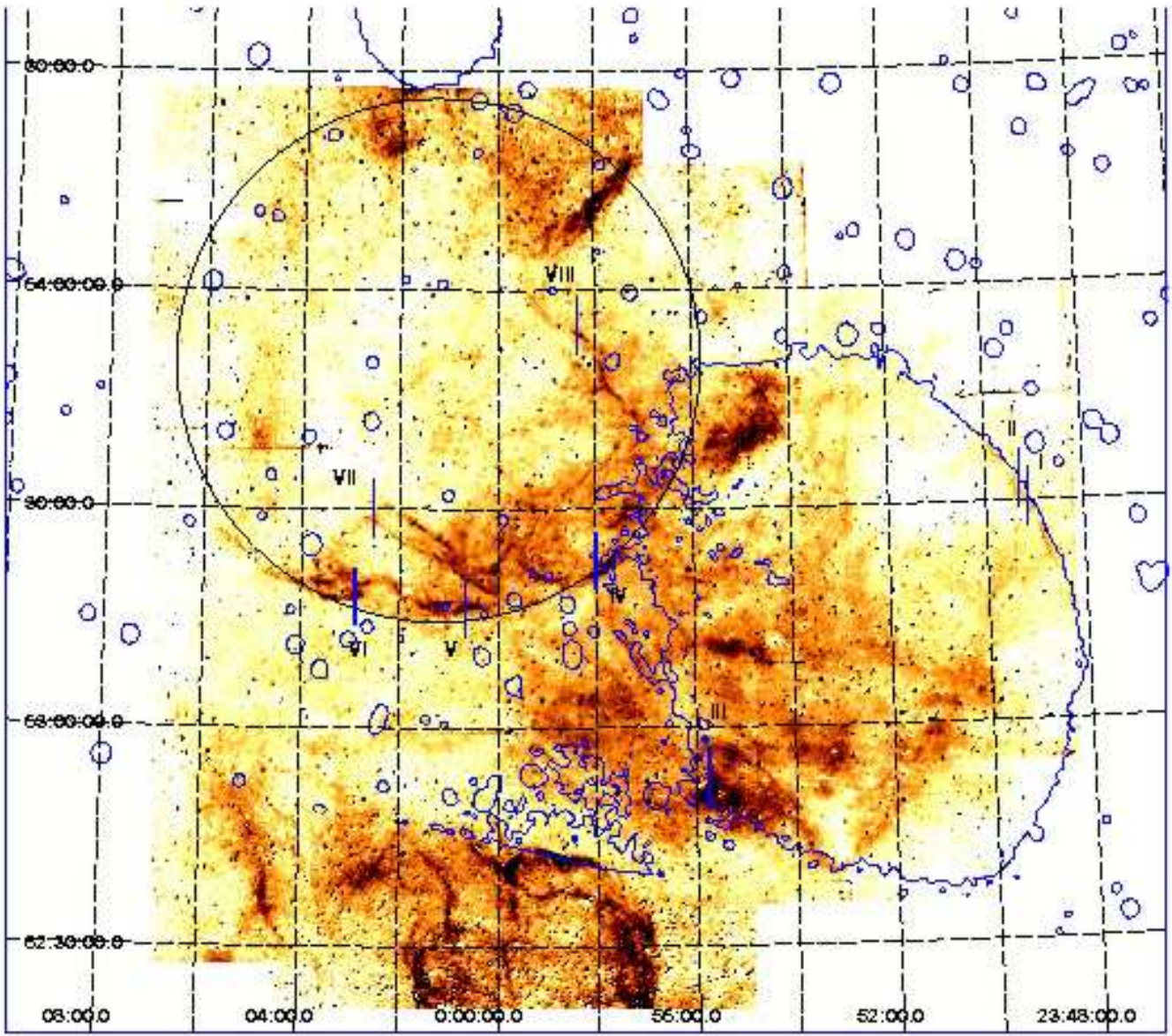}}
    \caption{
     Diffuse and patchy emission as well as new filamentary structures 
     are detected in the observed fields. The known supernova remnant
     \gsnr\ is located to the lower--right, where the 1400 MHz
     contour marks the remnant's actual location and size. 
     The circle seen to the upper--left part of the figure encompasses
     the new filamentary structures which may be part of a new candidate 
     remnant. The filaments seen in the middle--bottom  part of the figure
     constitute the upper portion of the well--known remnant CTB~1. 
     The long rectangles mark the slit projections on the sky and are
     numbered from I through VIII. The line segments seen 
     near over--exposed stars in this figure and the next figures 
     are due to the blooming effect.       
     The shadings run linearly from 0 to 32 $\times$ \flux.
     } 
     \label{fig01}
  \end{figure*}
  \begin {figure*}
  \resizebox{\hsize}{!}{\includegraphics{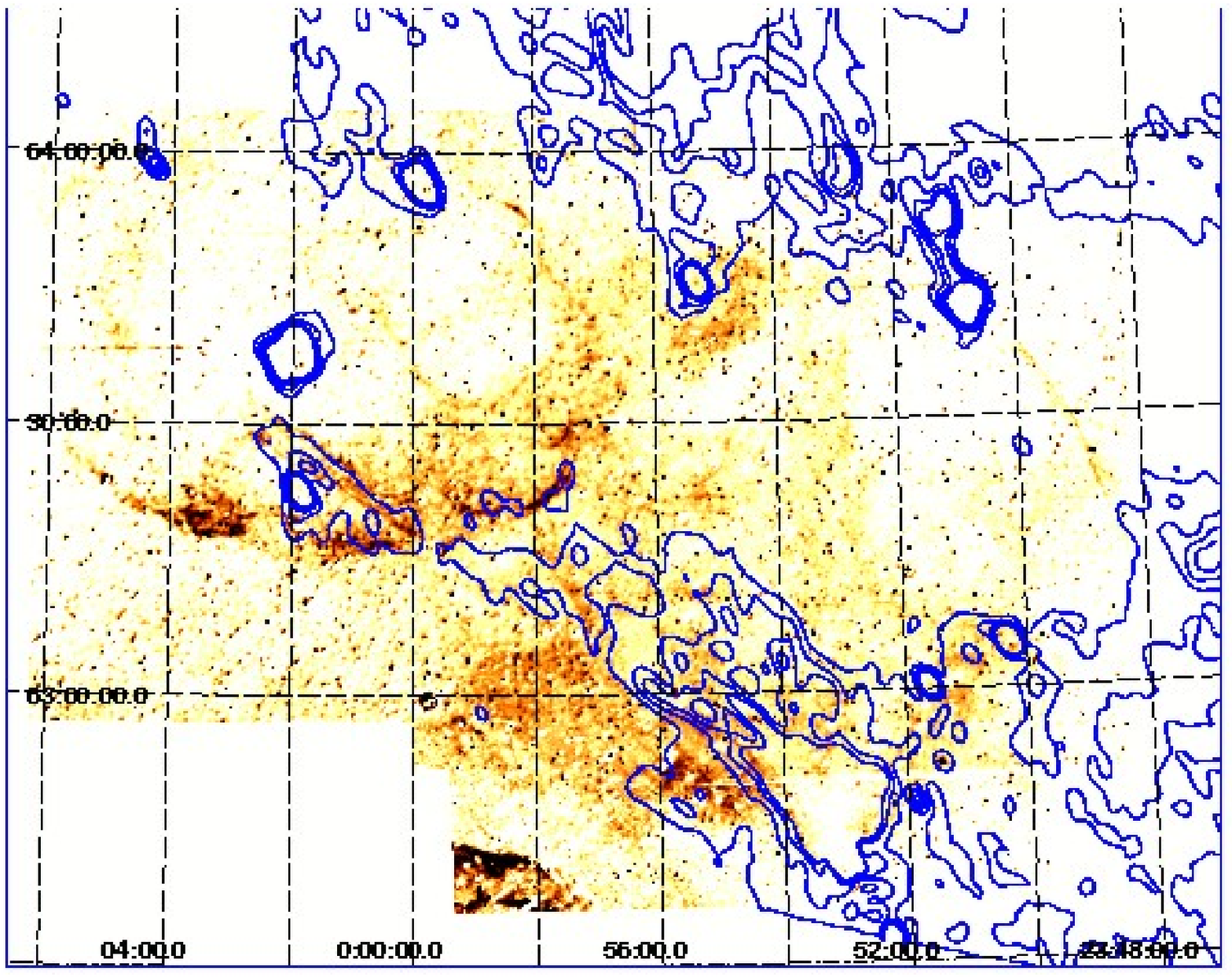}}
    \caption{The field of the remnant seen through the sulfur filter. 
    The shadings run linearly from 
    0 to 9 $\times$ \flux, while the 60$\mu$m contours scale linearly
    from 11 to 16.4 MJy sr $^{-1}$, every 1.8 MJy sr $^{-1}$. It may be possible that
    cool dust is associated with the two filaments extending to the north--east
    around \a\ $\simeq$ 00\h01\m30\s, and \dd\ $\simeq$ 63\degr26\arcmin; 
    short segments of infrared emission also overlap optical emission around 
    pos. IV (cf. Fig. \ref{fig01}). The intensity of these dust structures is
    typically 10$\sigma$ above the local IR background.
         } 
     \label{fig02}
  \end{figure*}

\subsection{Optical images}
The imagery was performed with the 0.3 m wide--field Schmidt--Cassegrain 
telescope at the Skinakas Observatory, Crete, Greece.
Multiple pointings were performed from August 9 to 13, 2002 in 
order to map the whole field of \gsnr\ as defined by its radio emission. 
The field to the east of \gsnr\ was observed on June 12, 14--15, and 20--22,
2004. 
In all cases, the telescope was equipped with a 1024 $\times$ 1024 
Thomson CCD providing a 70\arcmin\ $\times$ 70\arcmin\ field of view and an 
image scale of 4\arcsec\ per pixel. The filters isolating the 
\hnii, \sii, and \oiii\ emission lines were used in these observations. 
Details about the filter characteristics and the exposure times can be found 
in Table~\ref{filters}.
An astrometric solution was calculated for all image frames with the
aid of the HST Guide star catalogue (Lasker \et\ \cite{las99}) and all
frames were subsequently projected to a common origin on the sky for 
further operations. The equatorial coordinates quoted in this work refer to 
epoch J2000. 
\par
The data were reduced using standard IRAF and MIDAS routines. 
The available frames were bias subtracted and flat--field corrected using 
a series of well--exposed twilight flat--fields. 
The absolute flux calibration was performed through observations of a series of
spectrophotometric standard stars (HR5501, HR7596, HR7950, HR9087, 
HR718 and HR8634; Hamuy \et\ \cite{ham92}, \cite{ham94}). 
Flux measurements in the overlapping areas, between the August 2002 and 
June 2004 observations, show very good agreement (better than 10\%).
\subsection{Optical spectra}
Long slit spectra in the area of \gsnr\ were obtained on July 10  
and August 10, 11 and 13, 2002 with the 1.3 m Ritchey--Cretien telescope at 
Skinakas Observatory. 
Additional spectroscopy was performed on July 13 and 14, 2004 with
the same configuration as in the earlier observations consisting of a
1300 line mm$^{-1}$ grating and a 800 $\times$ 2000 SITe 
CCD covering the range of 4750 \AA\ -- 6815 \AA.
The slit width was 7\farcs7, the slit length was 7\arcmin.9 and it was always 
oriented in the south-north direction. The coordinates of the slit centers, the 
number of available spectra and the total exposure times are given in 
Table~\ref{spectra}. Sky background was always taken from the same frame as the 
source spectrum in areas of little or no measureable emission. 
Standard IRAF tasks were used to reduce the source and the auxiliary frames. 
Absolute flux calibration was performed through observations of the 
spectrophotometric standard stars HR7596, HR7950, and HR9087, HR718 
and HR8634 (Hamuy \et\ \cite{ham92}, \cite{ham94}).   
\subsection{The \hnii\ and \sii\ line images}
The field shown in Fig. \ref{fig01} encorporates several diffuse, patchy as 
well as filamentary structures. 
The known supernova remnant \gsnr\ is located to the lower--right where 
the single 1400 MHz contour marks the actual location and size of this object,
while the circle drawn to the upper--left encloses the new filamentary 
features that were discovered in this area. These new features may suggest 
the detection of a previously unknown remnant. Finally, 
part of the well--known supernova remnant CTB~1 can be seen in the 
middle--bottom area of the figure.
\par
The long, faint filament detected by Fesen \et\ (\cite{fes97}) is also 
present in our images and its typical intensity is a factor of 2-3 lower 
than the intensity of the structures found in the central to east areas 
of the remnant (pos. I in Fig. \ref{fig01}). The faint emission in the
south, around a declination of 62\degr39\arcmin, seems to be obscured by a dust
cloud giving rise to the ''V'' shape of the optical emission (see sect. 5). 
Stronger \hnii\ emission is detected in the eastern areas of \gsnr,
while the properties of the spectrum at position III (Fig. \ref{fig01}) 
may indicate emission from shock heated gas (Table \ref{sfluxes}; sect. 4). 
The new imaging observations were performed, mainly in the \hnii\ 
filter, in an attempt to trace the full spatial extent of the emission 
to the east of \gsnr. The diffuse emission in the very north edge of our 
field of view is part of \object{LBN 117.62$+$02.29} (Lynds \cite{lyn65}), 
an \HII\ nebulosity, while we were not able to
find a registered entry in the SIMBAD database for the structure around 
\a\ $\simeq$ 23\h58\m, \dd\ $\simeq$ 64\degr10\arcmin, extending for
$\sim$16\arcmin\ in the south--east to north--west direction.
\section{The imaging observations}
  \begin{table*}
        \caption[]{Relative line fluxes}
         \label{sfluxes}
         \begin{flushleft}
         \begin{tabular}{lllllll}
     \hline
 \noalign{\smallskip}
                      &Pos. Ia   & Pos. Ib & Pos. II &  Pos. IIIa &   Pos. IIIb &  Pos. IV         \cr
\hline
Line (\AA)             & F$^{\rm a,b}$ & F$^{\rm a,b}$  & F$^{\rm a,b}$ &F$^{\rm a,b}$ & F$^{\rm a,b}$           \cr
\hline
4861 \hbeta\            & --       & --      &--          &19 (3)   & 15 (2)   & 14 (2)               \cr
\hline
4959 \oiii          & --       & --      &--       	   &--       & --	& --               \cr
\hline
5007 \oiii          & --       & --      &--       	   &--       & --       & --               \cr
\hline  
6300 \oi            & 29 (3)   & 21 (5)  &--          &5 (3)    & 8 (3)    & 14 (3)              \cr
\hline
6364 \oi            &  --      &  --     & --      	   &    --   & --       & --              \cr
\hline
6548 \nii           & --       & 16 (4)  &7 (1)        &9 (5)    &6 (2)     & 7 (2)          \cr
\hline
6563 \ha\           & 100 (13) & 100 (23)&100 (10)     &100 (51) & 100 (35)	& 100 (27)         \cr
\hline
6584 \nii           & 43  (6)  & 50 (12) &40 (4)       &41 (21)  & 45 (16)	& 51 (13)           \cr
\hline
6716 \sii           & 47  (6)  & 52 (14) &36 (4)       &22 (12)  & 27 (11)  & 50 (14)           \cr
\hline
6731 \sii           & 32  (4)  & 39 (11) &33 (4)       &22 (12)  & 27 (11)  & 47 (14)           \cr
\hline 
\hline
Absolute \ha\ flux$^{\rm c}$ & 3.8 & 5.1      & 9.1        &14.1      & 13.9     & 11.5              \cr
\hline
\ha /\hbeta\            & --        & --       & --         & 5.3 (3)  & 6.6 (2)  & 6.9  (2)          \cr
\hline
\sii/\ha\               & 0.79 (7)  & 0.91 (14)& 0.69 (5)   & 0.44 (16) &0.54 (14)& 0.97 (16)         \cr
\hline 
I(6716)/I(6731)       & 1.47 (4)  & 1.32 (8) & 1.08 (3)   & 1.0 (8)  & 1.0 (8)  & 1.1 (10)          \cr
\hline 
\end{tabular}
\begin{tabular}{llllllll}
& & & & \cr
                      &Pos. V    & Pos. VI  & Pos. VII &  Pos. VIII        \cr
\hline
Line (\AA)             & F$^{\rm a,b}$ & F$^{\rm a,b}$  & F$^{\rm a,b}$ &F$^{\rm a,b}$         \cr
\hline
4861 \hbeta\            & 13 (4)   & 18 (6)    & 21 (3)    & 9 (2)                \cr
\hline
4959 \oiii         & --       & --        &--    	   &--                     \cr
\hline
5007 \oiii          & --       & 9 (4)     &--     	   &--                      \cr
\hline  
6300 \oi            & 17 (12)  & --        &---        & --                  \cr
\hline
6364 \oi            & 6 (4)    &  --       & --        & --              \cr
\hline
6548 \nii           & 11 (8)   & 15 (12)   & 13 (2)    & 10 (2)           \cr
\hline
6563 \ha\               & 100 (70) & 100 (76)  &100 (20)   &100 (19)          \cr
\hline
6584 \nii           & 46  (32)  & 48 (37)   &30 (6)    &42 (8)        \cr
\hline
6716 \sii           & 55  (42)  & 60 (50)   &51 (13)   &60 (12)          \cr
\hline
6731 \sii           & 38  (30)  & 43 (35)   &36 (9)    &42 (8)          \cr
\hline 
\hline
Absolute \ha\ flux$^{\rm c}$ & 16.2  & 16.6      & 8.6      &8.7                   \cr
\hline 
\ha /\hbeta\            & 7.9 (4)     & 5.7 (6)   & 4.8 (3)   & 11 (1)            \cr
\hline
\sii/\ha\               & 0.93 (41)   & 1.03 (47) & 0.87 (12) & 1.02 (11)          \cr
\hline 
I(6716)/I(6731)       & 1.44 (24)   & 1.40 (29) & 1.4 (7)   & 1.4 (7)          \cr
\hline 
\end{tabular}

\end{flushleft}
 ${\rm ^a}$ Fluxes uncorrected for interstellar extinction and relative to
 F(\ha)=100

${\rm ^b}$ Listed fluxes are a signal to noise weighted
average of the individual fluxes

$^{\rm c}$ In units of \flux\ 

${\rm }$ Numbers in parentheses represent the signal to noise ratio 
of the quoted fluxes \\
\\\end{table*}

%
\par
The overall morphology in the \sii\ filter (Fig. \ref{fig02}) is similar to 
that of the \hnii\ filter, however, some structures in the field appear 
stronger in this line than others. This is the 
first indication that we may have detected emission from shock heated gas. 
We take advantage of our flux calibrated images (see also Mavromatakis, 
Xilouris and Boumis \cite{mav04} for a description of the method) to 
estimate roughly the \sii/\ha\ ratio over the observed areas. 
The sulfur to \ha\ ratios are calculated only for those areas where the 
scatter of the measured fluxes, among the different nights of observation, is 
$\sim$ 15\% or less.
It is found that over the area of \gsnr\ most of the detected emission 
is likely to originate from photoionized gas with the exception of the known
filamentary emission in the west. Emission in area III can be asssociated 
with the
remnant but more observations are required to verify this. 
\par
The \sii/\ha\ ratios over the newly detected structures in the 
north--east strongly support the fact that the gas has been shock heated 
rather than photoionized. This conclusion is verified by the deep long 
slit spectra which offer more accurate measurements of the individual 
line fluxes (sect. 4).
No significant \oiii\ emission was detected and the 3$\sigma$ upper limit, 
over the area of \gsnr\ and the new fields, is $\sim$4$\cdot$\flux.
\section{The long slit spectra of \gsnr}
\begin {figure}
  \resizebox{\hsize}{!}{\includegraphics{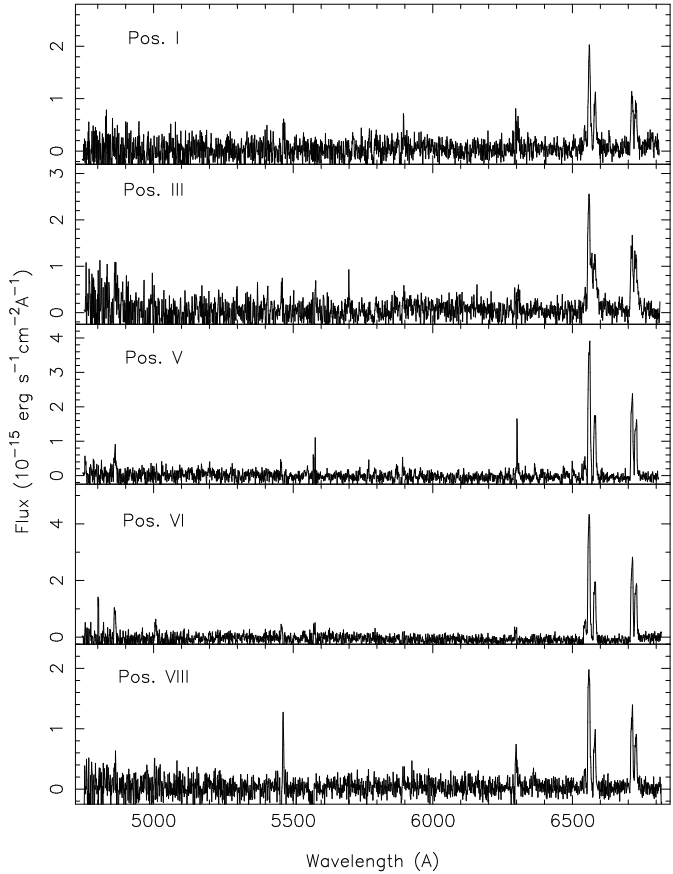}}
    \caption{Long slit spectra from individual frames extracted from the
    listed positions in the area of \gsnr\ and its surrounding area. 
    More details can be found in sect. 4 and Table \ref{sfluxes}.
         } 
     \label{fig03}
  \end{figure}

Deep long--slit spectra were taken in order to establish accurately  
the nature of the observed emission by measuring the strengths 
of the \ha\ and \sii\ emission lines. A number of spectra, extracted 
from individual data frames, are shown in Fig. \ref{fig03} and
the measured fluxes are given in Table \ref{sfluxes}.
\par
A set of spectra were obtained to the west of \gsnr\  (pos. I, II) 
and a \sii/\ha\ 
ratio $\sim$0.9 was measured (Table \ref{sfluxes}), identifying the optical 
emission as emission from shock heated gas (see also Fesen \et\ \cite{fes97}). 
Two different apertures were extracted from the spectrum obtained at pos. III.
The sulfur to \ha\ ratios are 0.4 and 0.5, close to the lower limit of values 
for supernova remnants and this does not allow us to unambiguously identify 
the origin of this emission as arising from shock heated gas. 
The origin of the emission remains an open question, although the \sii\ image 
(Fig. \ref{fig02}) shows enhanced emission in this area. 
We note here that the \sii/\ha\ ratio, in known SNRs, is typically greater 
than 0.4 and mainly above 0.5, (e.g. Fesen \et\ \cite{fes85};  
Raymond \et\ \cite{ray88}; Smith \et\ \cite{smi93}), while in \HII\ regions it 
is found below $\sim$0.35 and mostly around 0.2 (e.g. Hunter \et\ \cite{hun92}). 

\par
Five different positions were observed spectroscopically in the area 
of the new filaments.
The properties of the spectra at positions IV, V, VI and VII strongly
point to emission from shock heated gas (\sii/\ha\ ratios close to 1; 
Table \ref{sfluxes}). The filamentary nature of the newly discovered optical
radiation, as seen in the narrow band images, also supports 
this conclusion. Although, the low ionization lines are quite strong,
\oiii\ line emission at 5007 \AA\ is only detected at pos. VI 
(Table \ref{sfluxes} and Fig. \ref{fig03}) and its flux is consistent with the 
upper limit obtained from the imaging observations. Finally, a single 
spectrum was obtained in the north where another filamentary structure is 
detected (pos. VIII). The spectrum here suggests that the observed emission 
originates from shock heated gas. 
\par
In those cases where the sulfur lines are accurately measured, 
the implied electron densities are low, e.g. less than 180 \dens\ at the 
3$\sigma$ limit at pos. VI. However, if the sulfur lines are weak and thus, 
of lower significance, the electron densities can cover a wide range of 
values, e.g. from 30 to 1400 \dens\ at the 3$\sigma$ limit at pos. IV. 
%
\par
In the absence of reliable \hbeta\ measurements for \gsnr, we adopt the statistical
relation of Predehl and Schmitt (\cite{pre95}) to estimate the color excess. 
The total galactic N$_{\rm H}$ in the direction of \gsnr\ is 7.3$\cdot$10$^{21}$
\sdens, according to the {\it FTOOLS} command ``nh'' based on data from Dickey
and Lockmann (\cite{dic90}). However, the velocity resolved N$_{\rm H}$ maps of
Hartmann and Burton (\cite{har97}) allow us to estimate a column density of
$\sim$5$\cdot$10$^{21}$ \sdens, assuming that the remnant is indeed in the
Perseus Arm (Reich and Braunsfurth \cite{rei81}). 
The color excess is then 1.09, equivalent to a logarithmic interstellar 
extinction of 1.66. 
Sufficient \hbeta\ emission has been detected only at pos. VI, in the area
of the new filamentary structures. An interstellar extinction of 0.8 is 
calculated (\ha/\hbeta\ $=$ 5.6), which is equivalent to a 
color excess of 0.5.
\section{Observations in other wavelengths}
X--ray counterpart emission, above the sky background, has not been detected in 
the area of \gsnr\ or the new candidate remnant in the ROSAT All Sky survey.
A possibly extended X--ray source (\object{1RXS
J235257.3+625002}) is present in the south but it is not at all clear if 
it is related to \gsnr. 
Despite the failure to detect soft X--rays, HI kinematic data 
(Taylor \et\ \cite{tay03}) as well as high resolution infrared data (Cao \et\
\cite{cao97}; Kerton and Martin \cite{ker00}) are available and these could 
provide additional information about the properties of this remnant. 
\par
New HI kinematic data are available from the Canadian Galactic Plane Survey 
(CGPS; Taylor \et\ \cite{tay03}). 
In Fig. \ref{fig05} we have split the velocity 
interval from --70 \vel\ to --30 \vel\ into eight narrower channels and radio 
contours at 1400 MHz are shown on top of the HI emission. The depression of the 
HI emissivity shown in Fig. 6 of Reich 
and Braunsfurth (\cite{rei81}) is mainly present between 
--67.4 and --61.0 \vel. 
The HI images in the range of --45 to --36 \vel\ are interesting because 
intense HI emission is present just outside the radio boundaries in the north 
and south of the remnant. Weaker and diffuse emission appears to be projected 
over the west part of the remnant. The HI emission is significantly reduced 
over the north--east part of \gsnr\, where the radio emission is also weak with
respect to other areas of the remnant. If the reduced HI emission at the center
and the intense HI emission at the boundaries of \gsnr\ are due to its expanding
envelope then we can estimate distances in the range of 2.5 to 3.5 kpc (Clemens
\cite{cle85}). However, the actual distance depends also on the adopted galactic
rotation curve even in the case where a correlation is well established.
%

%
\par
We have also examined infrared (IR) data at 8.28 $\mu$m 
(MSX\footnote{The MSX data were retrieved from the NASA/IPAC
Infrared science archive at http://irsa.ipac.caltech.edu}), and 
12, 25, 60 and 100 \mm\ (IRAS). 
In Fig. \ref{fig02}, the 60 $\mu$m data are shown in the form of light, thin 
contours along with the 1400 MHz radio emission (dark, thick contours).
Intense dust emission in the south--east of \gsnr\ 
results in the ``V'' shape of the optical radiation.
A close--up of this area is shown in Fig. \ref{fig04}, where the 
8.28 $\mu$m emission displays a comet--like structure. The bulk of the IR 
emission is confined in an area less than $\sim$10\arcmin\ long, while the 
overall structure extends for $\sim$30\arcmin. 
The right plot in Fig. \ref{fig04} shows the HI emission in the velocity range
--45 to --41 \vel\ along with contours of the 60 $\mu$m emission. The intense
IR and HI emission seem to match spatially, while there is also good 
positional agreement in the two lanes of emission extending to the north--east.  
If the correlation proposed in Fig. \ref{fig04} is true, then it would imply 
that the optically emitting gas is further than $\sim$3 kpc away (Clemens
\cite{cle85}). 
\par
Further to the north-east in Fig. \ref{fig02}, it can be seen that IR 
emission traces the two filamentary structures present between a right 
ascension of 00\h00\m\ and 02\h00\m, just south of the 
declination of 63\degr30\arcmin. In addition, patchy IR emission at pos. IV
appears to match the optical emission at this location and a correlation 
could be possible. 
The intensity of the IR emission, of $\sim$11 MJy sr$^{-1}$, is 
10$\sigma$ above the neighbouring background level of 
$\sim$8 MJy sr$^{-1}$.

%
\section{Discussion}
The supernova remnant \gsnr\ was observed in major optical emission lines
and deep long slit spectra were acquired at a few positions.
The overall morphology is similar among the low ionization images (Fig.
\ref{fig01}, Fig. \ref{fig02}). 
New patchy and diffuse structures are detected in the area of \gsnr\  
but they do not seem related to it. There is an area of emission 
in the south--east which could be related to the remnant and this is the
emission detected around pos. III. One of the extracted apertures is 
characterized by \sii/\ha\ $\sim$ 0.5 and weak \oi 6300\AA\ emission, while the 
\sii\ image shows enhanced radiation in this area compared to neighbouring
positions. Its non--filamentary nature is not necessarily an indication of 
photoionized gas since remnants, like e.g. \object{W44}, display 
similar morphologies (Raymond \& Curiel \cite{ray95}; 
Mavromatakis \et\ \cite{mav03}).
The lower \sii/\ha\ ratio measured here, compared to the north--west, 
could be explained by a lower ionization state of the preshocked gas 
and/or by a stronger than usually assumed magnetic field 
(e.g. Raymond \cite{ray79}; Cox \& Raymond \cite{ray88}).
\par
The long slit spectra from the area of \gsnr\ set an even lower upper limit 
to any \oiii\ emission at the specified positions compared to the imagery. 
The \oiii\ flux  production depends mainly upon the shock velocity and the 
ionization state of the preshocked gas. Given the data in hand, 
we cannot determine whether slow shocks travel into ionized gas or whether 
faster shocks travel into neutral gas (Cox \& Raymond \cite{cox85}), but 
we can exclude moderate or fast shocks overtaking ionized gas. 
It is evident that the \oiii\ image data help us to generalize this 
conclusion and apply it to the whole area of the remnant.
%
  \begin {figure*}
  \resizebox{\hsize}{!}{\includegraphics{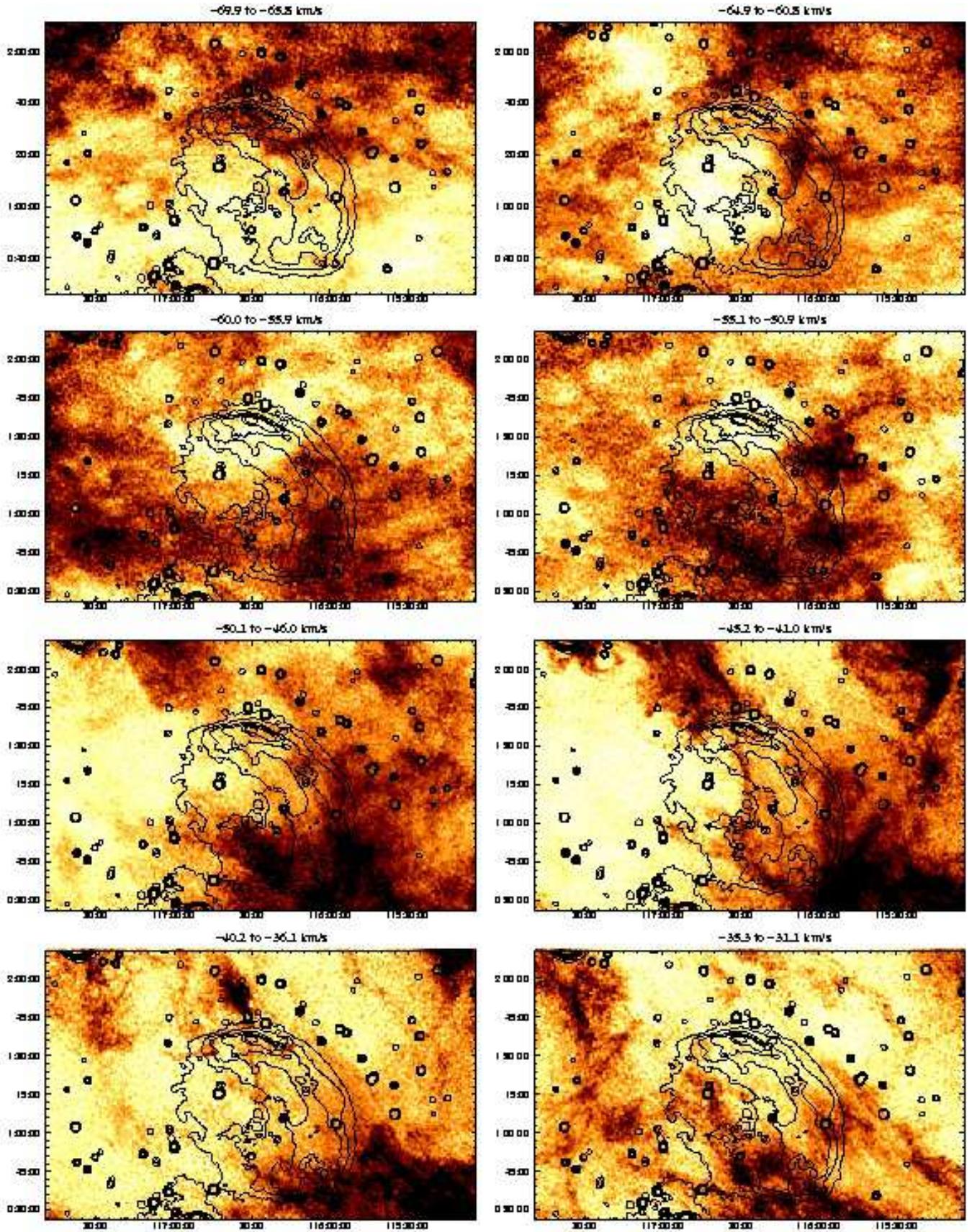}}
    \caption{The HI emission in the range of LSR velocities --70 to --31 \vel\ 
    has been split into eight narrower velocity intervals. 
    The radio contours at 1400 MHz 
    scale, every 0.22 K, from 5.7 to 6.8 K brightness temperature. The data are
    plotted in galactic coordinates and more details are given in sect. 5. 
         } 
     \label{fig05}
  \end{figure*}
%
  \begin {figure*}
  \resizebox{\hsize}{!}{\includegraphics{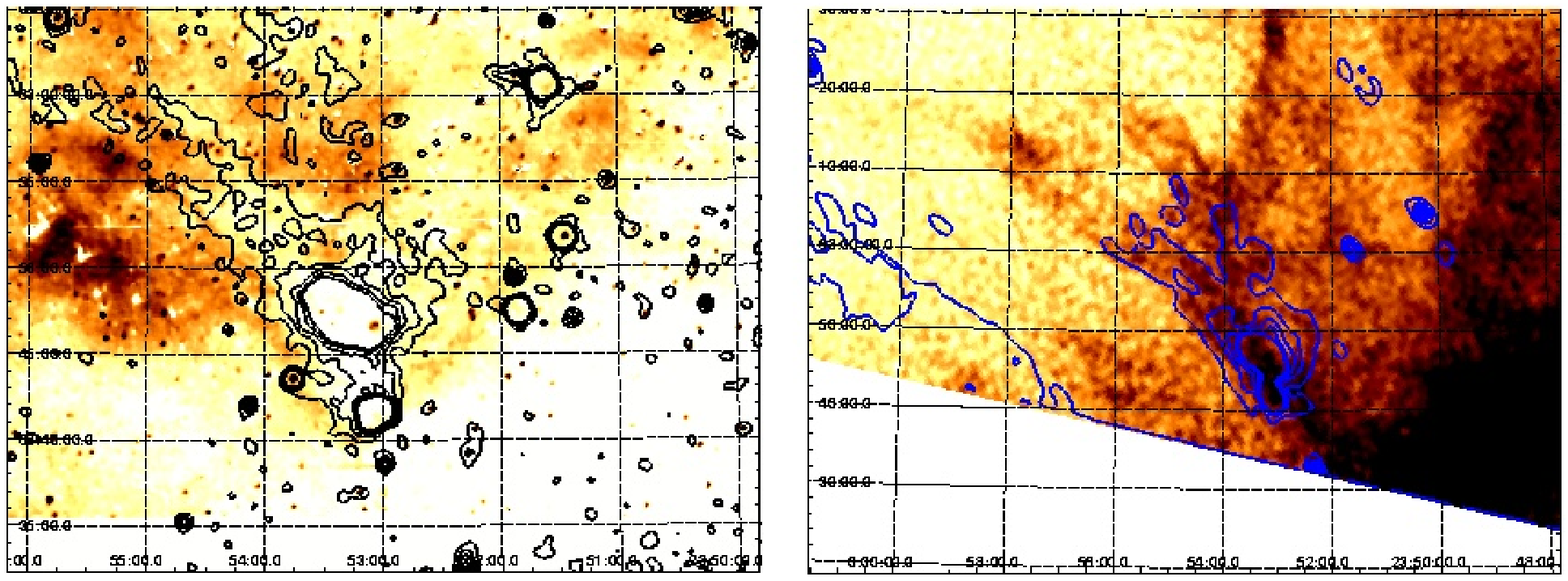}}
    \caption{ The left figure shows the optical emission in the south of \gsnr\
    along with high resolution MSX data at 8.28 \mm. The IR emission is nicely
    anticorrelated with the \hnii\ emission. Note that the central IR emission 
    is a factor of $\sim$15 stronger than the emission of the tail extending 
    to the  north--east. The right figure shows the HI emission in a wider 
    area and in the range of LSR velocities --45 to --41 \vel, while the 
    overlaid contours outline the 60 \mm\ IR data 
    (1\arcmin$\times$1\arcmin.7 resolution;  
    Cao \et\ \cite{cao97}) and is seen that the infrared emission also extends 
    to the north--east.
         } 
     \label{fig04}
  \end{figure*}

\par
If the radio and HI emission over \gsnr\ are indeed associated in the 
range of --45 to --36 \vel, then the distance would be in the range
of 2.5 to 3.5 kpc (sect. 5; Fig. \ref{fig05}). This distance is compatible
with that of 3.1 kpc calculated by Huang and Thaddeus (\cite{hua86}) using 
the $\Sigma$--D relation, while the longer distance given by Reich and
Braunsfurth (\cite{rei81}) may be due to the different galactic rotation
curve used and the wider range of the adopted HI velocities 
of --60 to --45 \vel. Radio emission at a level of 5.8 MJy sr$^{-1}$ 
best matches the filamentary optical emission in the north--west, 
resulting in an angular size of 79\arcmin\ $\times$ 51\arcmin. 
This is equivalent to a linear size of 
69 pc $\times$ 45 pc, at a distance of 3 kpc, placing
this object among the largest known supernova remnants. 
\gsnr\ is an interesting remnant to study, especially in X--rays, since its 
typical length scale is difficult to accomodate with ages of a few  
10$^4$ years and E51/n$_{\rm o}$$\sim$10 or less 
(E51 stands for the explosion 
energy in units of 10$^{51}$ erg and n$_{\rm o}$ is the initial density of 
the ambient medium).
\par
Several filaments have been discovered to the north--east 
of \gsnr. 
A number of these filamentary structures (e.g. pos. IV, V, VI) trace 
well the periphery of the 36\arcmin\ radius sphere whose boundary 
is drawn in Fig. \ref{fig01}. 
Other structures appear to be projected in the inner areas 
(e.g. pos. VII, VIII). Estimates of the \sii/\ha\ ratio based
on the images show that we are dealing with radiation coming from gas that has
been shocked rather than photoionized. Indeed, the long slit spectra at five
different positions clearly identify the emission as resulting from shock heated
gas. In order to obtain estimates of the preshock cloud density 
n$_{\rm c}$, we assume a shock velocity in the range of 70 to 120 \vel, 
given the apparent absence of \oiii, and find values of n$_{\rm c}$ around 5 
\dens\ or less 
(Fesen \& Kirshner \cite{fes80}). The \hbeta\ flux, although weak, is present 
in the majority of the spectra and this could imply that the
new structures are closer to us than \gsnr. Their morphology,  
spatial distribution on the sky and spectral signatures
lead us to propose that we have observed a new candidate remnant which we
designate as G~117.4$+$1.5. 
Furthermore, these data (optical and radio) do not support a
relation between this new candidate remnant and the known remnant \gsnr.
\par
In the area of the new candidate remnant, weak infrared emission at 
60 $\mu$m and 100 $\mu$m appears to match the optical 
emission around pos. IV and VII (Fig. \ref{fig02}), while at shorter IR 
wavelengths significant emission is not detected. The CGPS data at 408 and 1400 
MHz were examined for radio emission in the area of the new candidate remnant, 
but we were not able to 
identify any morphological features that would imply a correlation with the
optical emission. The failure to detect radio emission at these frequencies 
may lie in a number of reasons, like the magnetic field strength, the ambient 
medium density, shock conditions etc. (e.g. Blandford and Cowie \cite{bla82}; 
Pineault \et\ \cite{pin97}). 
However, this object may not be the only one where optical emission is 
detected but at the same time its radio emission is very weak or below 
detection.
A candidate remnant in Cygnus, near CTB~80, displays a similar morphology
in the optical and very weak radio emission (Mavromatakis \& Strom
\cite{mav02}). 
Another, very extended and evolved, candidate remnant in Pegasus shows highly
filamentary structures in the optical, while associated radio emission has not 
been detected (Boumis \et\ \cite{bou02}).
It may be possible that the candidate remnant G~117.4$+$1.5 
belongs to the same class of objects. It is clearly interesting to note that
Blandford \& Cowie (\cite{bla82}) had predicted the existence of ``radio quiet'' 
remnants. New observations of this region, 
in wavelengths other than optical and especially in radio, would be required 
to validate our suggestion. In addition these observations could determine the 
properties of such objects and explain why they have escaped detection by 
current radio surveys.  
\section{Conclusions}
The emission in the field of \gsnr\ is diffuse and patchy  and the majority of
the emission may not be related to the remnant. 
There is only one area in the central--east where the
detected emission could be related to the remnant. If true, 
it would suggest that \gsnr\ is more extended in the optical regime than 
previously realized. The distance of $\sim$3 kpc derived 
from HI kinematic data agrees fairly well with previous estimates and implies 
a very extended remnant.
\par
The low ionization images and the long slit spectra of an area to the
north--east of \gsnr\ show that the discovered emission most likely originates 
from shock heated gas. The morphology of the emission is filamentary, while
some of the filaments appear aligned along the periphery of a circle of 
36\arcmin\ radius. 
These filaments  are probably part of a previously unrecognized remnant 
designated as G~117.4$+$1.5. It is characterized 
by very low radio emission, below the detection
limit of current radio surveys.
\begin{acknowledgements}
The authors would like to thank the referee for stimulating suggestions and 
comments which helped to improve and clarify the scope of this work. 
Skinakas Observatory is a collaborative project of the University of
Crete, the Foundation for Research and Technology-Hellas and
the Max-Planck-Institut f\"ur Extraterrestrische Physik.
This research made use of data products from the Midcourse Space 
Experiment.  Processing of the data was funded by the Ballistic 
Missile Defense Organization with additional support from the NASA 
Office of Space Science. The data from the Canadian Galactic 
Plane Survey were obtained from the Canadian Astronomy Data Centre
(where the author is a guest user) which is operated by the Herzberg
Institute of Astrophysics of the National Research Council Canada.
\end{acknowledgements}
%

\vfill\eject
%
%
\end{document}